\documentstyle[12pt,epsf]{article}
\def\321{ SU(3)_C \otimes SU(2)_L \otimes U(1)_Y }
\def\41{ SU(4)\otimes U(1) }
\def\5{\underline{5}}
\def\bar5{\underline{\overline{5}}}
\def\24{\underline{24}}
\newcommand{\be}{\begin{equation}}
\newcommand{\ee}{\end{equation}}
\newcommand{\beqa}{\begin{eqnarray}}
\newcommand{\eeqa}{\end{eqnarray}}
\newcommand{\eqn}[1]{(\ref{#1})}

\newcommand{\barh}{\overline{H}}
\def\Tr{\hbox{{\rm Tr}}}
\begin{document}
\setlength{\unitlength}{1mm}
 
{\hfill hep-ph/9707405}

{\hfill DSF 35/97} 

{\hfill SISSA 95/97/EP} \vspace*{2cm}

\begin{center}
{\Large \bf Hybrid Inflation from Supersymmetric SU(5)}
\end{center}

\bigskip\bigskip
 
\begin{center}
{\bf L. Covi$^{1,2}$}, {\bf G. Mangano$^{3}$}, {\bf A. Masiero$^{1,4}$}, 
and {\bf G. Miele$^{3}$} 
\end{center}
 
\vspace{.5cm}

\noindent
$^{1}$ {\it International School for Advanced Studies, SISSA-ISAS,
Via Beirut 2/4, I-34014,Trieste, Italy }\\
$^{2}$ {\it INFN, Sezione di Trieste, Via A. Valerio 2, I-34127 Trieste, 
Italy}\\
$^{3}$ {\it Dipartimento di Fisica, Universit\'a di Napoli "Federico II", and
INFN, Sezione di Napoli, Mostra D'Oltremare Pad. 20, I-80125 Napoli, Italy}\\
$^{4}$ {\it Dipartimento di Fisica, Universit\'a di Perugia, and INFN,
Sezione di Perugia, Via Pascoli, I-06123 Perugia, Italy}
\bigskip\bigskip\bigskip
 
\begin{abstract}
A scheme of hybrid inflation is considered in the framework
of the minimal supersymmetric $SU(5)$ model with an extra singlet. 
The relevant role of the cubic term in the adjoint representation 
in the renormalizable superpotential is pointed
out in order to have a quite wide region of initial conditions
compatible with inflation efficiency, monopole density dilution and 
perturbations constraint.  
\end{abstract}
\vspace*{2cm}
 
\begin{center}
{\it PACS number(s):12.60.Jv; 98.80.Cq}
\end{center}
 
\newpage
\baselineskip=.8cm
Unified gauge theories certainly provide a natural framework
where to look for a proper scalar dynamics able to drive inflation.
The debate on the inflationary option to adopt to match the most
recent observations is still open \cite{old}-\cite{hyb}. A particularly
attractive scenario is hybrid inflation \cite{hyb}. In this class of
models, by using more than one scalar field, one 
naturally overcomes the unpleasant feature to have an incredibly 
small coupling constant, typical of single scalar field models \cite{hyb2}.
As far as the elementary particle physics is concerned, supersymmetric 
GUT models still remain one of the most appealing candidates for next step in
unification programme \cite{susy}, and naturally provide many scalar fields.
The simplest GUT example capable to unify all gauge interactions is
the supersymmetric $SU(5)$ model \cite{susysu5}.

There already exist some examples of simplified SUSY hybrid realizations
\cite{hyb2}. Here we intend to construct a hybrid inflationary scheme
based on a realistic SUSY $SU(5)$ model with an additional singlet superfield.
In particular we will show that it is possible to avoid a characteristic 
problem of such approaches, an excessive production of magnetic monopoles
at the end of inflation.

Let us consider the following globally supersymmetric renormalizable 
superpotential
\be
W = \mu^2 ~s - \alpha  ~s~ \Tr(\Phi^2)- \beta \Tr(\Phi^3)+
\gamma ~\barh \Phi
H + \delta ~\barh H~~~,
\label{1}
\ee
where $s$ is a complex gauge singlet field, $\Phi$ denotes a 
complex $\24$ irreducible representations (IRR) of $SU(5)$, and 
$H_a$ and $\barh_a$, with $a=1,..5$, stand for a $\5$ and 
$\bar5$ IRR's of $SU(5)$, respectively. Without loss of generality
we can choose $\mu^2,\alpha,\beta $ as positive real constants by a suitable
redefinition of complex fields.  
We can express $\Phi$ in components with respect to the $SU(5)$ adjoint 
basis\footnote{We normalize $T^i$ as $\Tr(T^i T^j) = \delta_{ij}/2$.}
$ \Phi = \phi_i T^i$. We will denote with the same letter the chiral superfields
or its scalar component depending on the context.
For simplicity we omit higher order terms in $s$. 

In terms of components, the superpotential becomes
\be
W = \mu^2 ~s - {\alpha \over 2}  ~s~ \Sigma_{i=1}^{24} \phi_i^2- 
{\beta \over 4} d_{ijk} \phi_i \phi_j \phi_k + 
\gamma  T^i_{ab}~\barh_a \phi_i 
H_b + \delta \barh_a H_a~~~,
\label{2}
\ee
where $d_{ijk} \equiv 2 \Tr(T^i\{T^j,T^k\})$.
Using Eq. \eqn{2} one gets the potential for the scalar components of the 
Higgs superfields $s$, $\Phi$, $H$ and $\barh$
\beqa
V\left(s,\Phi,H,\barh\right)  =  \left| \mu^2 - 
{\alpha \over 2} ~ \Sigma_{i=1}^{24} \phi_i^2\right|^2 
+ \Sigma_{i=1}^{24}
\left|\alpha s \phi_i + {3 \beta \over 4} d_{ijk}
\phi_j \phi_k - \gamma T^i_{ab} \overline{H}_a H_b \right|^2
\nonumber\\
+\Sigma_{b=1}^{5}\left(
\left| \gamma T^i_{ab} \phi_i \overline{H}_a + \delta \overline{H}_b 
\right|^2 + \left|\gamma T^i_{ba} \phi_i H_a + \delta
H_b \right|^2 \right) + (D-\mbox{terms})~~~.
\label{3}
\eeqa
Then the absolute minimum of the supersymmetric potential results to be
\beqa
\Sigma_{i=1}^{24} (\phi^0_i)^2 = { 2 \mu^2 \over \alpha}~~~,
~~H^0_a~=~\barh^0_a=0~~~,~~~
s^0 = -{3 \beta \over 8 \mu^2} d_{ijk}\phi^0_i\phi^0_j\phi^0_k~~~.
\label{4}
\eeqa
We can always use an $SU(5)$ transformation to put the v.e.v. matrix
$\Phi^0$ into diagonal form, $\phi^0_i \neq 0$ only along the four Cartan 
diagonal generators.
The request $ \phi^0_j = \phi^{0*}_j $ for such components both
ensures vanishing D-terms and verifies the first condition of Eq. \eqn{4}
where the r.h.s. is positive.
The superscript $0$ reminds that the background configurations
correspond to the absolute minimum. 

To realize the $SU(5)$ breaking into $\321$, all $\phi_i$ will vanish 
apart of $\phi_8 \equiv v/\sqrt{2}$ (hypercharge component) and 
$s\equiv\sigma/\sqrt{2}$ ($v$ and $\sigma$ two real scalar fields).
Thus, by virtue of \eqn{4} one has two possible solutions:
\be
v^0 = \pm {2 \mu \over \sqrt{\alpha}}~~~\mbox{and}~~~\sigma^0=\pm 
{3 \beta \mu \over \sqrt{60} \alpha^{3/2}}~~~\mbox{accordingly}.
\label{5}
\ee
The parameters $\mu $ and $ \alpha $ of our superpotential
are therefore connected to the Higgs expectation value.

As usual to have doublet-triplet splitting in $H$ and $\barh$, 
a fine tuning on the potential parameters is required. 
Let us denote by $h_3$ ($\overline{h}_{3}$) and $h_2$ ($\overline{h}_{3}$)
the above triplet and doublet components of $H$ ($\barh$). By using the 
expression \eqn{3} for the potential $V$, evaluated at the absolute 
supersymmetric minimum \eqn{4}, one gets the following mass terms
\be
\left(\delta + { \gamma v^0 \over \sqrt{30}}\right)^2
\left(h_3^{\dag} h_3 + \overline{h}_3 \overline{h}_3^{\dag}\right)
+\left(\delta - { 3 \gamma v^0 \over 2 \sqrt{30}}\right)^2
\left(h_2^{\dag} h_2 + \overline{h}_2 \overline{h}_2^{\dag}\right)
~~,\label{6}
\ee
where $ v^0 $ has one of the values in Eq. \eqn{5}.
Thus, in order to have the doublets massless one has to impose the
fine tuning condition
\be
\delta = {3 \gamma v^0 \over 2 \sqrt{30}}~~~.
\label{7}
\ee 

Let us now consider the inflationary scenario emerging in the framework of
our model,
with chaotic initial conditions. According to this picture,
the initial values for $\Phi$ and $s$, emerging from the quantum cosmological
period, are arbitrary, and in general do not coincide with the supersymmetric
absolute minima \eqn{5}. For simplicity we will assume however 
that the direction 
of this fields is already $\321$ invariant, and it remains so during all
the evolution. This is a simplifying assumption since one should
consider a general $SU(5)$ breaking and follow the dynamics of all
the independent fields without fixing a preferred direction. At
the moment our goal is to see if it is possible to have a viable 
inflationary scenario in the context of a simplified SUSY $SU(5)$ model. 

 The consistency of our assumption requires at least that the 
chosen direction 
should be locally stable with respect to small perturbations in any other 
direction, as for example the one invariant under $\41$. 
This is actually the case, since we have verified that all the above 
perturbations have positive mass square in the region 
$v$--$\sigma$ we are going to study (see later Fig. 2).
 
In terms of the real scalar fields $v$ and $\sigma$, the classical 
potential of Eq.\eqn{3} becomes
\be
V\left(\sigma,v\right) = \left(\mu^2 - {\alpha \over 4} v^2 \right)^2+ 
{\alpha^2 \over 4} v^2 \left(\sigma - \xi v \right)^2~~~, 
\label{8}
\ee
where $\xi \equiv 3 \beta/(2 \sqrt{60} \alpha)$. 

Let us for convenience rewrite the potential in terms of adimensional
variables rescaling all the fields with respect to $ v^0 $ and the scalar 
potential with respect to the mass scale $\mu$:
\beqa
y &=& {\sqrt{\alpha} \over 2 \mu} v~~~, ~~~~~~~~
x= {\sqrt{\alpha} \over 2 \mu} \sigma~~~, \\
\widetilde{V}&=& { V \over \mu^4} = (1-y^2)^2 + 4 y^2 (x-\xi y)^2~~~.
\eeqa

In Fig. 1 we show the potential for a particular value of $\xi$. We can see
that it is characterized by two valleys, one corresponding to $ y = 0 $
($SU(5)$ symmetric configuration) and the other with $ x= \xi y $. 
It is interesting to notice that although the second valley displays for
large values of $ x $ a higher energy density with respect to the first one, 
it is exactly the second pointing toward the supersymmetric minimum.

In terms of these quantities, the early universe scalar dynamics reads
\beqa
\ddot{y} &+& 3 \widetilde{H}\dot{y} + {\delta\widetilde{V}\over\delta y} = 
0~~~,\\
\ddot{x} &+& 3 \widetilde{H}\dot{x} + {\delta\widetilde{V}\over\delta x} = 
0~~~,\\
\dot{N} &=& \widetilde{H} = K \left[{\dot{y}^2\over 2} + {\dot{x}^2\over 2}
+ \widetilde{V} \right]^{1/2},
\label{9}
\eeqa 
where $ K $ is defined as $ \sqrt{32\pi\mu^2 /(3\alpha M_{Pl}^2)} $ and 
$N$ denotes the e-foldings. 
In the previous equations the dot indicates the derivative
with respect to the dimensionless time $ \tau = t \mu\sqrt\alpha/2 $
and also the Hubble parameter $\widetilde{H}$ is properly rescaled.
Notice that all the dynamics depends only on the initial conditions and on
the parameters $K$ and $\xi$.

We can proceed to a numerical integration of Eq. \eqn{9} to obtain the
classical dynamics of the fields starting with arbitrary initial conditions
$x_i$ and $y_i$.
Depending on the initial values we have observed that 
two possible dynamics take place:
in one case the fields enter the $SU(5)$ symmetric valley and the 
symmetry is restored during the inflationary period to be broken later
at the end of inflation. We have then production of monopoles that cannot
be diluted by a sufficiently large e-foldings.
On the contrary, for other initial conditions, the symmetry is no
more restored and the fields slowly descend toward the supersymmetric minimum
along the valley $x= \xi y$. In this case any topological defects production
took place before inflation and can be efficiently diluted.
In Fig. 2 we show some trajectories in the $x-y$ plane for particular 
initial conditions. The lines starting at A and B correspond to large 
initial conditions and, as can be seen clearly, both follow the second valley
of the potential up to the supersymmetric minimum. They yield a large
number of e-foldings, 150 and 40 respectively. On the contrary line C 
describes a situation in which the fields fall in the $ SU(5)$
symmetric valley until practically the end of inflation. 
In Fig. 2 are well indicated 
for positive $x$ and $y$ (the case of both negative $x$ and $y$ is 
completely analogous) the two regions of initial conditions corresponding
to the two described behaviour.
For $x_i > 0$ and $y_i < 0$ or viceversa
the dynamics proceeds towards symmetry restoration and is therefore not
favorable.

The field behaviour for lines A and B in Fig. 2 can be understood 
qualitatively looking at the dynamics in the directions singled out by 
the valley (one orthogonal and the other along the bottom of the valley); 
defining:
\be
z = { x - \xi y \over \sqrt{1+\xi^2}}~~~,~~~~
w = { y + \xi x \over \sqrt{1+\xi^2}}~~~,
\ee
we have that the gradient of the potential is given by 
\beqa
{\delta \widetilde{V} \over \delta z} &=& 8 z ( w-\xi z)^2-8\xi z^2 ( w-\xi z)
+ {4 \xi \over 1+\xi^2} (w-\xi z)(1 -{(w-\xi z)^2 \over 1+\xi^2})~~~,
\label{10} \\
{\delta \widetilde{V} \over \delta w} &=& 8 z^2 ( w-\xi z)
- {4 \over 1+\xi^2} (w-\xi z)(1-{(w-\xi z)^2 \over 1+\xi^2})~~~.
\label{11}
\eeqa
It is easily seen from these expressions that the force along the
$z$ direction is greater than the one along $w$ for $\xi$ larger than
unity and
all choices of initial conditions corresponding to a negative $z_i$ and 
large $w_i$ (i.e. large $ x_i \simeq y_i $), since Eq. \eqn{11} is suppressed 
by a factor $1/\xi$ with respect to part of Eq. \eqn{10}. 
The r.h.s. of Eq. \eqn{10} is large and negative and pushes $z$ to zero;
correspondingly we will have first an essentially one-dimensional motion 
along $z$ and then an analogous behavior along $w$ for $z = 0$.

The parameter $\xi$ affects only the first part of the dynamics; it must be
sufficiently large to have well separated valleys defined by $ x = \xi y$ and
$ y = 0$. In this case $\xi$ rules the velocity of approaching from
the initial point the bottom of the valley. Fundamental in stopping 
the fields in the valley is the strength of
the frictional term proportional to $K$.
The second part of dynamics, now independent of $\xi$, is only 
affected by $K$. Again the friction term must be large enough to
slow down the fields so that they do not overcame the $y=0$ barrier
separating the minima at $y= \pm 1$,
and start small oscillations around the supersymmetric minimum.  
Interestingly the numerical analysis shows that an acceptable behaviour
is obtained for $ K \geq 7\cdot 10^{-2}$. 

Whenever $K$ is chosen in this range, the initial conditions for $x$ and
$y$ cannot be chosen too small as indicated by the shaded region in Fig. 2.
Apart from this bound, starting in the allowed region, to small
initial conditions correspond small e-folding.

For the $SU(5)$ asymmetric dynamics, we have an initial inflationary period
while $z$ is approaching zero and then a second stage of inflation during 
the motion along the valley. 
For very special initial condition both phases of inflation are
necessary to give the $60$ e-folds able to solve the smoothness 
problem, and in this case the computation of the scalar density
perturbation is more involved \cite{st87}. However the second stage of 
inflation is generally long enough to produce an e-fold number of the 
order of $100$ and then it is possible to compute the scalar density 
perturbation neglecting the previous history.
The dynamics along $w$ fairly satisfies slow roll conditions;
by considering the potential along the bottom of the valley as a 
function of $w$, one can easily estimate the slow-roll parameters
$\epsilon $ and $\eta$ \cite{ckll}
\beqa
\epsilon &=& {1 \over 6 K^2} \left( {\widetilde{V}'
 \over
\widetilde{V}} \right)^2 \leq 0.3\cdot 10^2 \left( {\widetilde{V}' \over
\widetilde{V}} \right)^2~~~, \label{epsilon} \\
\eta &=& {1 \over 3 K^2} {\widetilde{V}'' \over
\widetilde{V} } \leq 0.6\cdot 10^2 {\widetilde{V}'' \over
\widetilde{V}}~~~, \label{eta}
\eeqa
where the primes denote the derivatives with respect to the $w$ variable
and the bounds are given considering the bound on $K$ due to the dynamics.
Both the quantities depending on the potential are smaller 
than $1$ for $ w \geq 30 $, i.e. the slow roll conditions break down
only when the fields approach the supersymmetric minimum $ w = \sqrt{1+\xi^2}$.
 In particular the value $w_e$ at which the inflation ends is 
given by $\max \{\epsilon,|\eta| \}=1$.

Since along the valley the dynamics is effectively one-dimensional and
satisfies the slow roll conditions, we can also easily evaluate the cosmic 
background radiation quadrupole anisotropy \cite{pert}.
The limit on scalar density fluctuations gives
\beqa
\left( {\Delta T \over T} \right)_Q \simeq \sqrt{32 \pi \over 45}
\left| { V^{3/2} \over V' M^3_{Pl}} \right|_{w_Q} = 
 { 2 K \over \sqrt{15}} \left( {\mu\over M_{Pl}} \right)^2 
\left( {\widetilde{V}^{3/2} \over \widetilde{V}'} \right)_{w_Q} 
\leq 7\cdot 10^{-6}~~~,
\eeqa
where the subscript $w_Q$ indicates the value of $w$ as the scale, which 
evolved to the present horizon size, crossed out the horizon during
inflation. The e-folding evaluated from $w_Q$ till the end of the 
inflation, $w=w_e$, is of the order of $N(w_Q) \simeq 60$.

From Eq.s (\ref{epsilon}) and (\ref{eta}) we easily get for $w_e$
\be
w_e^2= 1+ \xi^2 + {2 \over K^2} + {2 \over K^2} 
\sqrt{1+{2\over 3} (1+\xi^2)K^2}~~~,
\label{we}
\ee
and for the e-folding $N(w_Q)$ 
\be
N(w_Q) = \frac 38 K^2 \left[ w_Q^2 -w_e^2 - (1+ \xi^2) \log \left( 
{w_Q^2 \over w_e^2} \right) \right]~~~,
\label{nw}
\ee 
which implicitely provides $w_Q$ as function of $K$ and $\xi$ only.

The background radiation quadrupole anistropy can therefore 
be expressed as
\be
\left( {\Delta T \over T} \right)_Q \simeq {3 \over 64 \pi\sqrt{15}}
\alpha K^3 {(1+\xi^2 - w_Q^2)^2 \over (1+\xi^2) w_Q}~~~.
\ee
which using the COBE experimental value turns into a value for $\alpha$ 
depending on $K$ and $\xi$.
Actually, using Eq.s (\ref{we}) and (\ref{nw}) one finds that $\alpha$ 
is a very slowly
varying function, and in particular $\alpha \simeq 1 \div 5 \cdot 10^{-5}$ for 
$.1 \leq K \leq 1$ and $1 \leq \xi \leq 10$.

The bound on $K$ can be also expressed as a lower limit on the mass
$m_V$ of heavy vector bosons
\be
{m_V \over M_{Pl}} = g_5 ~\sqrt{ {5 \over 32 \pi} } K \geq g_5~1.6~ 10^{-2}~~~,
\ee
with $g_5$ the $SU(5)$ gauge coupling constant, which is fairly compatible
with present estimates which take into account threshold effects \cite{t1}--
\cite{t3}.

Using the allowed values for $\alpha$ one can also evaluate the 
masses of 
colour octet and weak triplet components of the adjoint Higgs. From
the expression of the potential, expanded around the absolute SUSY minimum
one gets
\be
{m_\Phi \over M_{Pl}} = \sqrt{15 \over 32} {\beta v^0 \over M_{Pl}} 
= \sqrt{75 \over 16\pi} \xi \alpha K  
\ee
which, for $\xi \sim 1 \div 10$ are of the order of $10^{13} \div 
10^{15}$~GeV.
Indeed such values are lower 
than the GUT scale, but consistent with those required in 
\cite{t1}--\cite{t4}  . Our scenario beautifully agrees with such case and
displays a high value for leptoquark gauge boson mass, of the order of 
$10^{17}$~GeV, which is consistent with the unification of coupling 
prediction only for
low values of the Higgs octet and triplet masses $m_\Phi$.

A final remark on the spectral index for scalar perturbations. 
In the slow--roll limit one has
\be
n(w_Q) = 1 + 2 \eta(w_Q) - 6 \epsilon(w_Q)
\ee
which using Eq.s (\ref{epsilon}), (\ref{eta}) and (\ref{nw}) gives
$n(w_Q) \sim 0.96$ for $.1 \leq K \leq 1$, a result which is compatible
with present determinations.

In this letter we have shown how a hybrid inflationary scenario can be
successfully realized in the framework of a realistic model based on 
$SU(5)$ supersymmetric
gauge theory, which represents one of the best candidates to describe
fundamental interactions up to very large scales beyond the Standard Model.
Actually starting from a renormalizable potential, the customary
Higgs content of MSSM $SU(5)$ with only an extra singlet, as first considered 
in a toy model in \cite{hyb2}, it is possible to obtain large e-fold number
and compatibility with scalar perturbation limits coming from COBE data. In 
particular the production of monopoles at the symmetry breaking occurs before
the inflationary stage for a quite wide range of initial conditions for the
scalar fields, so their density is strongly diluted in these cases. Finally
no fine tuning of coupling constant is needed in order to fit with cosmological
constraints.

We thank G. Dvali, G. Lazarides and Q. Shafi for useful discussions. The work 
of A.M. was partially supported by the EU contract ERBFMRX CT96 0090.

\newpage

\begin{figure}[Fig.1]
\centerline{\epsfbox{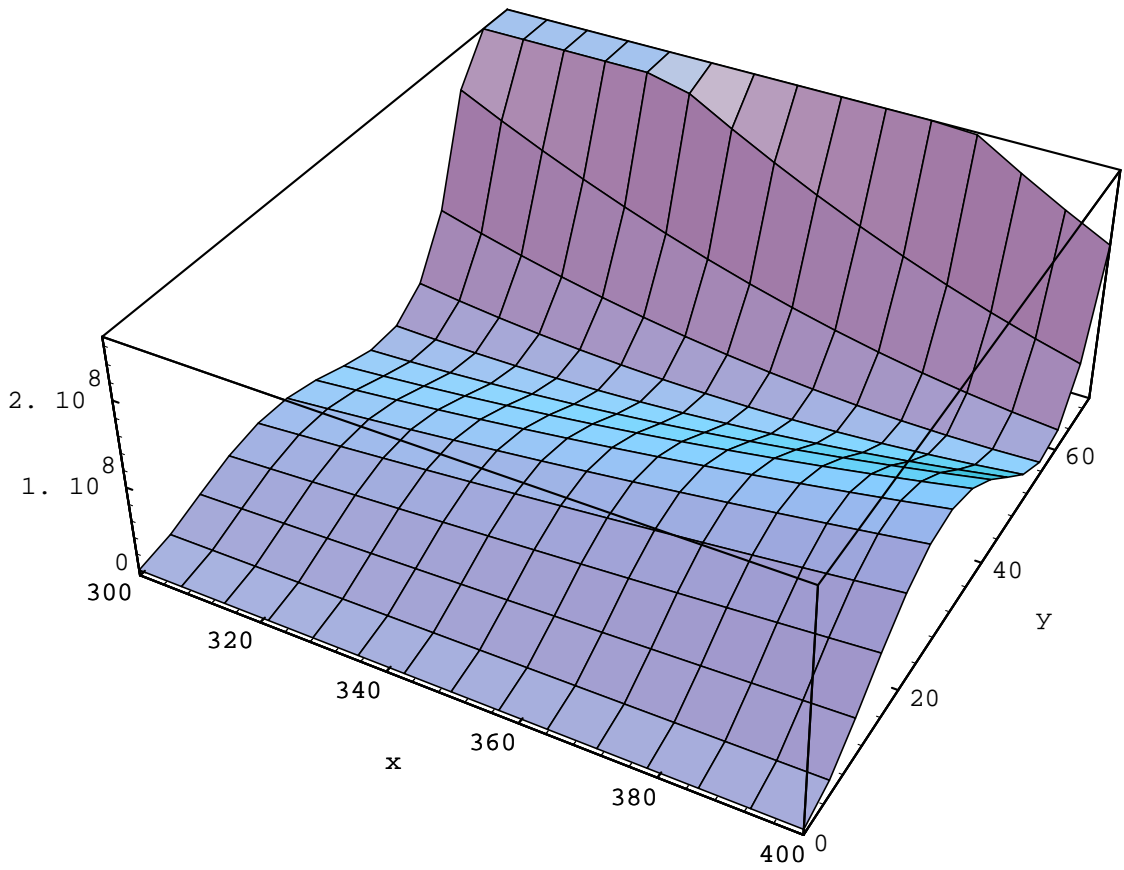}}

\vspace{1.5truecm}

\caption[]{The dimensionless potential $\widetilde{V}$ for $ \xi = 7 $ is
here shown.} 
\end{figure} 

\newpage

\begin{figure}[Fig.2]
\epsfysize=18cm
\epsfxsize=14cm
\epsffile{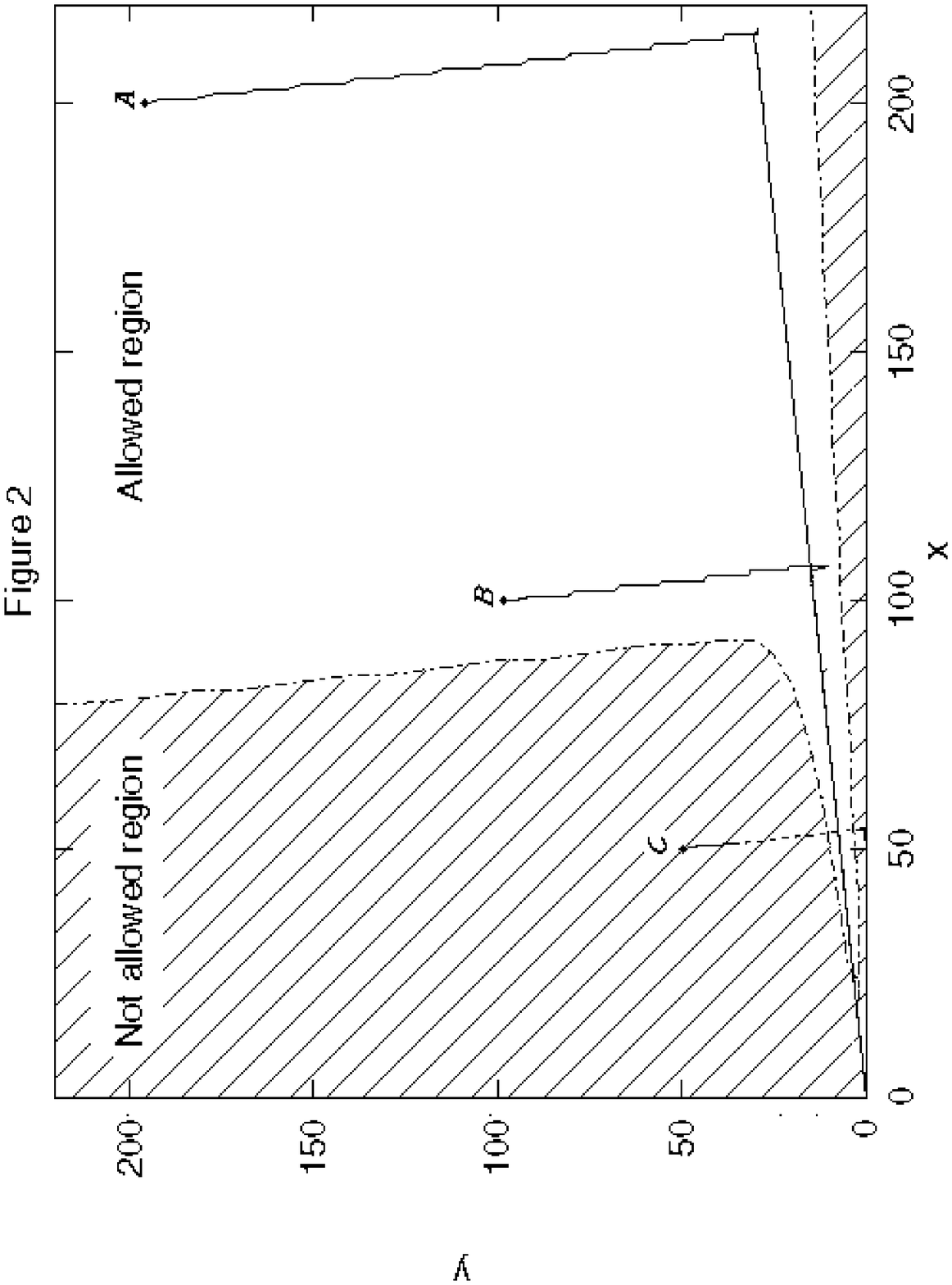}

\vspace{1.5truecm}

\caption[]{The curves A,B and C start from different initial conditions 
$(x_i,y_i)$, respectively, from the right, (200,200), (100,100), 
(50,50). The first two correspond to $N=150$ and $40$, respectively. 
The third one falls into the symmetric $SU(5)$ potential valley.
The shaded region contains the initial conditions leading to $SU(5)$ symmetry 
restoration during the inflationary dynamics.} 
\end{figure}

\end{document}